# Categorization of surface polarity from a crystallographic approach

Yoyo Hinuma[1*], Yu Kumagai[2], Fumiyasu Oba[2,3], and Isao Tanaka[1,4,5]

[1]Department of Materials Science and Engineering, Kyoto University, Sakyo, Kyoto 606-8501, Japan

[2]Materials Research Center for Element Strategy, Tokyo Institute of Technology, Yokohama 226-8503, Japan

[3]Materials and Structures Laboratory, Tokyo Institute of Technology, Yokohama 226-8503, Japan

[4]Elements Strategy Initiative for Structural Materials, Kyoto University, Sakyo, Kyoto 606-8501, Japan

[5]Nanostructures Research Laboratory, Japan Fine Ceramics Center, Nagoya 456-8587, Japan

E-mail * yoyo.hinuma@gmail.com

**Abstract**

With *ab initio* codes that employ three-dimensional periodic boundary conditions, the slab-and-vacuum model has proven invaluable for the derivation of energetic, atomistic, and electronic properties of materials. Within this approach, polar and nonpolar slabs require different levels of treatment, as any polar instability must be compensated on a case-by-case basis in the former. This article proposes an efficient algorithm based on isometries to identify whether a slab with the given surface orientation would be intrinsically polar, and if not, to obtain information on where to cleave the bulk crystal to obtain a stoichiometric nonpolar slab and whether reconstruction is necessary to generate a stoichiometric slab that is not polar.

Keywords: Polar surface, Nonpolar surface, Isometry, Slab-and-vacuum model, High-throughput calculations

## 1. Introduction

Understanding how atoms and molecules interact with the surface is a crucial problem in catalysis, photocatalysis, gas sensing, and crystal growth. The first step towards the modeling of such scenarios is the identification of which surface a material will preferentially adopt as many properties depend strongly on this. For example, surface orientations define electronic properties, such as the work function (WF), ionization potential (IP), and electron affinity (EA). These quantities not only provide the Fermi level and band positions with respect to vacuum and adsorbate levels at surfaces but also allow for prediction of the Schottky barrier heights of metal-semiconductor interfaces, band offsets of semiconductor heterointerfaces, and doping limits [1-4].

First principles calculations are an excellent tool to complement experimental investigations of crystal surfaces. Atom positions are usually explicitly provided in these calculations. This allows handling of both existing and hypothetical crystals, modeling of arbitrary reconstruction, and positioning of adsorbates at desired positions. Excluding cluster calculations, surfaces are usually simulated using a slab-and-vacuum model under three-dimensional periodic boundary conditions. Infinitely extending two-dimensional thin films are separated from their images by finite vacuum in this setup. Being capable of automatically generating a supercell containing a slab with a given arbitrary surface orientation is a necessity in the current age of "high-throughput" calculations [5, 6]. Gale and Rohl [7] discuss the potential complexity in surface creation. A surface can be specified by the Miller indices of the plane, which defines the orientation of the bulk cleavage, and the shift, or displacement of the plane relative to the origin. Complex cases can have many shifts leading to distinct surfaces in a given plane. Furthermore, in the case of a dipolar surface, the dipole must be removed, for instance by movement of atoms between the surfaces of a slab. The GULP code [7] suggests the use of the GDIS code [8], which is a visualization program for display and manipulation of isolated molecules and periodic structures, to specify surfaces by Miller indices, search valid shifts, and manipulate geometries. Sun and Ceder [9] provide a simple algorithm to find two basis vectors that span a given surface orientation and another basis vector that is maximally orthogonal to these two basis vectors. This has been implemented as a Matlab subroutine. Together with another Matlab subroutine they designed to transfer coordinate triplets of atoms when converting basis vectors from one to the other, their methodology allows relatively easy generation of a supercell with two basis vectors spanning a given surface orientation for any arbitrary crystal.

Unfortunately, the important problem of the polarity of the surface is not discussed by either Gale and Rohl [7] or Sun and Ceder [9]. Polar instability arises in polar surfaces where the macroscopic dipole moment perpendicular to the surface plane diverges when considered as a function of system thickness [10, 11]. A compensating electric field is necessary to resolve the polar instability, for instance, through (1) intrinsic surface charge modification by partial filling of electronic states, (2) intrinsic or extrinsic modification of the surface region composition, or (3) extrinsic adsorption of charged foreign species. Multiple mechanisms may take place at the same time, and the active mechanism(s) may depend on the bulk electronic structure, bulk polarization, and external conditions. Defects, which frequently appear as a compensation mechanism, modify the reactivity of the surface. For this reason, polar and nonpolar surfaces have to be handled differently in calculations. Using a model with the right compensating mechanism is essential in polar surfaces, however the compensating mechanism, which is surface-dependent, can be difficult to identify, as is the case for the wurtzite ZnO (0001) and (000$\bar{1}$) surfaces ($c$-plane) [12-15].

The exact cancellation of the macroscopic dipole is non-trivial with finite number of defects in a finite cell, and the large supercells required to accommodate a sufficiently appropriate defect concentration can be expensive to calculate. Examples of approaches to resolve the polar instability include the introduction of a planar dipole layer in the middle of the vacuum region [16], the addition of a compensating ramp-shaped potential in the vacuum region that cancels the artificial field together with an energy correction term [17], and the modification of the Coulomb interaction such that the Coulomb interaction at a point $\mathbf{r}'$ from a charge at point $\mathbf{r}$ is proportional to $|\mathbf{r}'-\mathbf{r}|^{-1}$ below a certain cutoff distance and zero otherwise [18]. In any case, calculations of polar surfaces require a level of treatment that varies on an individual basis, and are, therefore, not well-suited for high-throughput calculations. In contrast, nonpolar surfaces are amenable to the derivation of relevant properties because there is no need to consider the compensating mechanism. For instance, the valence band offset at an interface between semiconductors can be reasonably predicted using the IPs of nonpolar surfaces as long as the chemical bonding of two materials constituting the interface are alike and the interface is not metallic [4]. Although calculations of polar and nonpolar surfaces are both important, there is clearly a need to detect whether a given surface can be nonpolar and to be able to generate a nonpolar slab-and-vacuum model of the surface, especially when considering systematic high-throughput calculations of many surfaces.

Tasker's categorization of surfaces into three distinct types [19] is widely used to classify the polarity of ionic compound surfaces. Planes in a Tasker's Type 1 surface are neutral with both anions and cations whereas those in a Tasker's Type 2 surface are charged and arranged symmetrically such that there is no dipole moment perpendicular to the unit cell. Gale and Rohl [7] separate Type 2 surfaces into Type 2a where "the anions and cations comprising the layers are not coplanar, but which allows for some surface cuts to split the layers in such a way as to produce no dipole" and type 2b that is similar to type 2a but some ions at the top must be moved to the bottom to remove the dipole. A Tasker's Type 3 surface is charged with a perpendicular dipole moment. Goniakowski *et al*. [11] employ the concept of a dipole-free bulk unit cell, which is a bulk unit cell that may involve incomplete layers and does not have a dipole moment along a given direction. The frozen bulk termination is polar if the surface cannot be obtained by simply piling up of dipole-free bulk unit cells and nonpolar if at least one dipole-free bulk unit cell exists that leaves the surface region empty. A previous work by Goniakowski and Noguera [20] define the weakly polar surface as a surface such as the $SrTiO_3$ (100) surface, where "charge redistribution required for the cancellation of the macroscopic electrostatic field does not induce as strong a modification of the electronic structure as that predicted on truly polar surfaces like MgO (111)". Stengel [21] uses Wannier ion charges, which are derived from Wannier orbitals belonging to each ion and are typically formal charges, to determine the dipole-free bulk unit cell.

The objective of this paper is to propose a method from a crystallographic approach that derives a set of basis vectors of a primitive cell where two of the basis vectors span an arbitrary orientation in an arbitrary crystal. Polar and nonpolar surfaces are defined from a crystallographic point of view by using isometries, and nonpolar surfaces are further categorized into three types. A procedure is outlined that identifies where to cleave the bulk crystal to obtain a nonpolar surface and to create a slab-and-vacuum model supercell that has a given minimum slab and vacuum thickness. The proposed algorithm will facilitate automatic categorization and creation of nonpolar surfaces, which would lead to efficient high-throughput surface calculations including surface energy, WF, IP, and EA calculations.

## 2. Categorization of surface polarity

We propose in this paper a new crystallographic categorization of surface polarity. One important feature is that information on the spatial charge is unnecessary, hence there is no need to analyze whether each layer of atoms is neutral, investigate what the nature of each bond is, or assign charge densities to each layer of atoms. All surfaces are either *polar* or *nonpolar*. A surface is polar when it is impossible to cut out a slab that looks identical when viewed from either direction normal to the surface (no identical termination).

Nonpolar surfaces are further categorized into three types. A surface is "*nonpolar type A*" if the surface is not polar and each layer of atoms is stoichiometric. This is a stricter requirement than the Tasker type 1 surface that could have charge neutrality based on formal charge in every layer, such as in the example of $SrTiO_3$ (001) discussed later in this paragraph. Remaining surfaces can be categorized into "*nonpolar type B*" and "*nonpolar type C*" surfaces. The surface is the former if the boundaries of the dipole-free bulk unit cell, in other words, the repeat unit with no dipole moment perpendicular to the surface, lies between layers of atoms. The surface is the latter if the boundaries must lie on layers of atoms. A nonpolar type C slab cannot, by definition, be simultaneously nonpolar and stoichiometric when simply cleaved from bulk. However, a nonpolar and stoichiometric slab can be obtained by reconstruction of the surface, for instance, by removing half of the atoms on the topmost layer on both sides or octopolar reconstruction [22]. Fig. 1 shows prototypes of the four categories as well as issues arising in high-throughput calculations. All nonpolar type A surfaces, including the rocksalt (100) surface, are Tasker type 1 surfaces because stoichiometric layers are, by definition, charge neutral. Nonpolar type B surfaces, such as the fluorite (111) surface, are generally Tasker type 2 surfaces. One situation where a nonpolar type B surface may be regarded as a Tasker 1 surface is when the repeat unit is the following three layers, all with neutral formal charge, that are positioned evenly apart: (a layer of species A), (a layer of cations B and anions C), and (another layer of species A). Nonpolar type C surfaces are mostly categorized as Tasker type 3 surfaces, and a representative example is the rocksalt (111) surface. This specific surface typically undergoes octopolar reconstruction [22] where 3/4 of atoms in the topmost layer and 1/4 of atoms in the next layer are removed. Some nonpolar type C surfaces such as the cubic perovskite $SrTiO_3$ (001) surface, which is terminated on each side by either a SrO or TiO$_2$ plane, may be considered nonpolar Tasker type 1 based on the Wannier charge approach [21]. Polar surfaces according to the crystallographic definition are generally Tasker type 3 surfaces and a typical example is the wurtzite (0001) surface. However, a polar surface according to our definition could be considered a Tasker type 1 surface in some cases. If we take perovskite SrTiO$_3$ and move each TiO$_2$ layer by a fixed distance in the same direction along the *c*-axis without moving the SrO layers, this symmetry breaking changes the polarity of the (001) surface from nonpolar type C to polar. However, as all SrO and TiO$_2$ planes are neutral based on formal charge, the surface can still be regarded as a Tasker type 1 surface. In short, barring some exceptions, nonpolar type A surfaces are Tasker type 1 surfaces, nonpolar type B surfaces are Tasker type 2 surfaces, and nonpolar type C and polar surfaces are Tasker type 3 surfaces. We note that it is possible, as Tasker pointed out using the fluorite (111) surface as an example, to choose a termination such that a surface is polar (type 3) even though it is also possible to cleave a nonpolar (type 2) surface [19]. We attempt to choose a termination, if possible, resulting in a nonpolar surface over a polar surface and a nonpolar type B surface over a nonpolar type C surface. This is because the energy required to compensate the macroscopic dipole moment is typically large, thus it is natural to choose a nonpolar termination if available. Moreover, the surface-dependent reconstruction mechanism must be identified when investigating a nonpolar type C surface, so these surfaces are not suited for high-throughput calculations.

Although handled using a supercell under three-dimensional periodic conditions, slabs intrinsically have two-dimensional translational symmetry because the presence of the surface breaks the translational symmetry perpendicular to the slab. The symmetry of a slab can be described using the 80 affine layer group types described in the International Tables of Crystallography E (ITE) [23]. However, derivation of all symmetry operations of the slab to obtain the full layer group type is not necessary in the scope of this study. Instead, we desire to know if a slab with a given surface orientation must be polar because isometries of a certain form are lacking and, if not, whether reconstruction of the surface is necessary to obtain a stoichiometric slab. The Scanning Tables in Chapter 6 of the ITE [23] cannot be directly used to obtain the symmetry of the slab because a slab has thickness but a section plane does not.

## 3. Notations

We investigate a slab of a crystal with the $(hkl)$ surface in this work. Indices $i = \{1,2,3\}$, variables $m$ and $n$ are positive integers, $\boldsymbol{I}$ is the identity matrix, and the floor and ceiling functions are denoted as $\lfloor x \rfloor$ and $\lceil x \rceil$, respectively.

*Basis vectors* are expressed using column vectors and are those of a cell under three-dimensional periodic boundary conditions. Any point $x$ can be represented by a column vector of *coordinate triplets* as $\boldsymbol{x}^{\mathrm{T}} = (x_1, x_2, x_3)$. The position of this point $x$ in a supercell with basis vectors $(\mathbf{a}_1, \mathbf{b}_1, \mathbf{c}_1)$ is denoted by $\boldsymbol{X} = (\mathbf{a}_1, \mathbf{b}_1, \mathbf{c}_1)\boldsymbol{x}$, and this point is in the supercell when $0 \leq x_i < 1$ for all $i$. A *transformation matrix* converts one set of basis vectors into another. For instance, a transformation matrix $\boldsymbol{M}$ converts basis vectors from $(\mathbf{a}_1, \mathbf{b}_1, \mathbf{c}_1)$ to $(\mathbf{a}_2, \mathbf{b}_2, \mathbf{c}_2)$ as

$$(\mathbf{a}_2, \mathbf{b}_2, \mathbf{c}_2) = (\mathbf{a}_1, \mathbf{b}_1, \mathbf{c}_1)\boldsymbol{M} = (\mathbf{a}_1, \mathbf{b}_1, \mathbf{c}_1)\begin{pmatrix} M_{11} & M_{12} & M_{13} \\ M_{21} & M_{22} & M_{23} \\ M_{31} & M_{32} & M_{33} \end{pmatrix}. \qquad (1)$$

The *conventional unit cell*, as defined according to Ref. [24], is the starting point in this work and can be obtained using, for instance, the spglib code [25]. Its basis vectors are denoted as $(\mathbf{a}, \mathbf{b}, \mathbf{c})$. The "*out-of-plane vector*" of the $(hkl)$ surface, $\mathbf{c}_{\mathrm{OP}}$, is defined as $\mathbf{c}_{\mathrm{OP}} = h\mathbf{a} + k\mathbf{b} + l\mathbf{c}$, and an "*in-plane vector*" $h'\mathbf{a} + k'\mathbf{b} + l'\mathbf{c}$ must satisfy $(h', k', l') \cdot (h, k, l) = 0$. An in-plane vector may or may not be perpendicular to the out-of-plane vector.

An "$(hkl)$ *primitive cell*" is defined as a primitive cell with basis vectors $(\mathbf{a}_\mathrm{P}, \mathbf{b}_\mathrm{P}, \mathbf{c}_{1\mathrm{P}})$ where $\mathbf{a}_\mathrm{P}$ and $\mathbf{b}_\mathrm{P}$ are in-plane vectors. An "$(hkl)$ *n-supercell*" is defined as an $1 \times 1 \times n$ supercell of the $(hkl)$ primitive cell and has basis vectors $(\mathbf{a}_\mathrm{P}, \mathbf{b}_\mathrm{P}, \mathbf{c}_{n\mathrm{P}})$. Different $(hkl)$ *n*-supercells are related by $\mathbf{c}_{n\mathrm{P}} = n\mathbf{c}_{1\mathrm{P}}$, and the $(hkl)$ 1-supercell is equivalent to the same as the $(hkl)$ primitive cell.

A slab with the $(hkl)$ surface can be obtained by simply removing all atoms from an $(hkl)$ $n$-supercell where the coordinate $x_3$ is not within the range $z_{n-} = z_{nc} - z_{nt}/2 \leq x_3 \leq z_{nc} + z_{nt}/2 = z_{n+}$. Here, $z_{nc}$ is defined as the "*slab center*" along the out-of-plane direction and $z_{nt}$ as the "*slab thickness*". The symbols $z_{n-}$ and $z_{n+}$ are used to denote the coordinate $x_3$ at the "*lower and upper slab boundaries*", respectively. Fig. 2(a) shows the (001) primitive cell (1-supercell) of WAl$_2$ (space group type $P6_422$, number 181) [26] that contains three layers of WAl$_2$. A slab with boundaries $(z_{1-}, z_{1+}) = (0, 2/3)$ is shown in Fig. 2(b), where we immediately find that $z_{1c} = 1/3$ and $z_{1t} = 2/3$.

An *isometry* [27], which is also called a motion or isometric mapping, is an instruction assigning a unique "image" point $\tilde{X}$ to each point $X$ in point space while all distances are kept invariant. An isometry can be represented using matrix formulation as

$$\begin{pmatrix} \tilde{x}_1 \\ \tilde{x}_2 \\ \tilde{x}_3 \end{pmatrix} = \begin{pmatrix} W_{11} & W_{12} & W_{13} \\ W_{21} & W_{22} & W_{23} \\ W_{31} & W_{32} & W_{33} \end{pmatrix} \begin{pmatrix} x_1 \\ x_2 \\ x_3 \end{pmatrix} + \begin{pmatrix} w_1 \\ w_2 \\ w_3 \end{pmatrix}. \qquad (2)$$

This isometry is also denoted as $\tilde{x} = Wx + w$ or $\tilde{x} = (W, w)x$, where $W$ is the *matrix part* and $w$ is the column *vector part*. This study uses the symbol # to denote matrix and vector part elements in an isometry within the semi-open interval $[0,1)$ (includes 0 but excludes 1). All # symbols are not necessarily the same number when there are multiple appearances of # in one isometry.

## 4. Derivation of the $(hkl)$ primitive cell

Isometries are used in this work to construct slabs with a given orientation and to identify the polarity of the slab. It is possible, based on the approach by Sun and Ceder [9], to build a slab with a given orientation without performing a symmetry search. However, analysis of the symmetry is inevitable to confirm that a slab is primitive as reduction methods based on lattice parameters only [28-30] cannot reduce the volume of the cell when the cell is not primitive, for example when there is centering. In contrast, finding a primitive cell is relatively trivial using isometries. An infinite number of isometries exist because, for an arbitrary lattice translation vector $t$, $(W, w+t)$ is an isometry if $(W, w)$ is an isometry. However, a primitive cell has exactly one isometry with the form $(I, w)$ where $w^T = (\#,\#,\#)$, that is, $w^T = (0,0,0)$. Symmetry search software such as the spglib code [25] are available to obtain all isometries with $w^T = (\#,\#,\#)$. Furthermore, investigation of the symmetry is necessary to identify the polarity of the slab. A nonpolar slab must have one or more of appropriate symmetry elements, which are inversion center, two-fold rotation or screw parallel to the surface, or a mirror or glide plane parallel to the surface. All these symmetry elements can be described using an isometry of a single specific form as discussed in Section 5.

There is a need to keep track of lattice points within the algorithm outlined below. The lattice can be handled with low computational cost using a hypothetical "*empty*" cell where a virtual atom exists on every lattice point. This empty cell is convenient because lattice vectors can be expressed as coordinate triplets of virtual atoms. The first step is to construct the empty cell that has the same basis vectors $(\mathbf{a},\mathbf{b},\mathbf{c})$ as the initial unit cell, which is the conventional cell in this work. A virtual atom is simply placed at coordinate triplet $w$ if $(I, w)$ is an isometry of the original crystal.

The next step is to find any supercell with basis vectors $(\mathbf{a}',\mathbf{b}',\mathbf{c}')$ where two basis vectors are in-plane vectors of the (*hkl*) surface and $\mathbf{c}'$ is chosen to be $\mathbf{c}' \times \mathbf{c}_{OP} = 0$. The number of zeros in the set $\{h,k,l\}$ may be 2, 1, or 0, and the transformation matrix $M'$ that transforms basis vectors as $(\mathbf{a}',\mathbf{b}',\mathbf{c}') = (\mathbf{a},\mathbf{b},\mathbf{c})M'$ is derived for each case as in Table 1. If $\det(\mathbf{a}',\mathbf{b}',\mathbf{c}') < 0$, the basis vectors are retaken by using $-\mathbf{b}'$ instead of $\mathbf{b}'$, that is,

$$M' \begin{pmatrix} 1 & 0 & 0 \\ 0 & -1 & 0 \\ 0 & 0 & 1 \end{pmatrix}$$

is used instead of $M'$.

The following process reduces the in-plane basis vectors to find the smallest supercell with basis vectors $(\mathbf{a}_P, \mathbf{b}_P, \mathbf{c}_S)$ where $\mathbf{a}_S$ and $\mathbf{b}_S$ are in-plane vectors and $\mathbf{c}_S \times \mathbf{c}_{OP} = 0$. This is possible through identification of vector parts of the following three isometries with $\mathbf{W} = \mathbf{I}$, which is carried by finding these three coordinate triplets of the virtual atom in the "empty" cell with basis vectors $(\mathbf{a}', \mathbf{b}', \mathbf{c}')$: (1) $\mathbf{w}_1^T = (w_{11}, 0, 0)$ with smallest $w_{11}$ between $0 < w_{11} \leq 1$, (2) $\mathbf{w}_2^T = (w_{21}, w_{22}, 0)$ with smallest $w_{22}$ possible between $0 < w_{22} \leq 1$ and $w_{21}$ between $0 \leq w_{21} < w_{11}$ that makes $(\mathbf{I}, \mathbf{w}_2)$ an isometry, and (3) $\mathbf{w}_3^T = (0, 0, w_{33})$ with smallest $w_{33}$ between $0 < w_{33} \leq 1$. The basis vectors are transformed by $(\mathbf{a}_S, \mathbf{b}_S, \mathbf{c}_S) = (\mathbf{a}', \mathbf{b}', \mathbf{c}')(\mathbf{w}_1, \mathbf{w}_2, \mathbf{w}_3)$. Gaussian lattice reduction is applied to $\mathbf{a}_S$ and $\mathbf{b}_S$ (Appendix A) to obtain a new set of basis vectors $\mathbf{a}_P$ and $\mathbf{b}_P$. If $\det(\mathbf{a}_P, \mathbf{b}_P, \mathbf{c}_S) < 0$, $-\mathbf{b}_P$ is used instead of $\mathbf{b}_P$. This supercell with basis vectors $(\mathbf{a}_P, \mathbf{b}_P, \mathbf{c}_S)$ is convenient when visualizing the (*hkl*) surface. Gram-Schmidt orthogonalization cannot be used instead of Gaussian lattice reduction because, in general, $\mathbf{a}_S \cdot \mathbf{b}_S \neq 0$. One subtle point is that $\mathbf{w}_2^T$ has the form $(w_{21}, w_{22}, 0)$ instead of $(0, w_{22}, 0)$. This is because finding lattice points, or virtual atom positions, one-dimensionally along $\mathbf{a}'$ and $\mathbf{b}'$ is not always sufficient in looking for a set of basis vectors that is primitive. Instead, lattice points in the two-dimensional plane spanned by $\mathbf{a}'$ and $\mathbf{b}'$ must be identified.

The final step is to obtain basis vectors of the (*hkl*) primitive cell, $(\mathbf{a}_P, \mathbf{b}_P, \mathbf{c}_{1P})$, from $(\mathbf{a}_P, \mathbf{b}_P, \mathbf{c}_S)$. The isometry $(\mathbf{I}, \mathbf{w}_4)$ in the supercell with basis vectors $(\mathbf{a}_P, \mathbf{b}_P, \mathbf{c}_S)$ is identified where $\mathbf{w}_4^T = (w_{41}, w_{42}, w_{43})$ and $0 \leq w_{41} < 1$, $0 \leq w_{42} < 1$, and $w_{43}$ is the smallest between $0 < w_{43} \leq 1$. We define $\mathbf{c}_{1P}$ as $\mathbf{c}_{1P} = (\mathbf{a}_P, \mathbf{b}_P, \mathbf{c}_S)\mathbf{w}_4$; if $w_{43} = 1$ then $\mathbf{c}_P = \mathbf{c}_S$.

Fig. 3 shows basis vectors $(\mathbf{a}, \mathbf{b}, \mathbf{c})$, $(\mathbf{a}_S, \mathbf{b}_S, \mathbf{c}_S)$, and $(\mathbf{a}_P, \mathbf{b}_P, \mathbf{c}_{1P})$ for the (100) surface of a face-centered cubic lattice. Here, $\mathbf{w}_1^T = (w_{11}, 0, 0) = (1, 0, 0)$,

$w_2{}^\text{T} = (w_{21}, w_{22}, 0) = (1/2, 1/2, 0)$ , $w_3{}^\text{T} = (0, 0, w_{33}) = (0, 0, 1)$ , and $w_4{}^\text{T} = (w_{41}, w_{42}, w_{43}) = (1/2, 1/2, 1/2)$ are used in the derivation of $(\mathbf{a}_\text{S}, \mathbf{b}_\text{S}, \mathbf{c}_\text{S})$ and $(\mathbf{a}_\text{P}, \mathbf{b}_\text{P}, \mathbf{c}_\text{1P})$. Table 2 shows $(hkl)$ primitive cell basis vectors for a number of surfaces in simple cubic, face-centered cubic, and body-centered cubic cells. The transformation matrix from the conventional cell to the $(hkl)$ primitive cell cannot be uniquely defined in other Bravais lattices because the transformation matrix depends on relations between basis vector lengths and interaxial angles, that is, axial ratios.

## 5. Identification of stoichiometric nonpolar slab type and boundaries

Fig. 4 is a flowchart of how the polarity type is determined in this section. The first step is to find out whether a nonpolar surface can be obtained or not. First, identification of potential centers of nonpolar slabs ("*potential slab centers*", $z_{nc}$) in an $(hkl)$ *n*-supercell is carried out. An isometry that provides information on a potential slab center must have the form

$$(\mathbf{W}, \mathbf{w}) = \begin{pmatrix} \# & \# & \# & \# \\ \# & \# & \# & \# \\ 0 & 0 & -1 & w_3 \end{pmatrix}. \quad (3)$$

When applying the isometry in equation 3 as $\tilde{\mathbf{x}} = \mathbf{W}\mathbf{x} + \mathbf{w}$, an atom with coordinate $x_3$ is mapped to $\tilde{x}_3 = w_3 - x_3$ after necessary in-plane rotation and/or translation. Therefore, $z_{nc} = w_3/2$ can be regarded as a potential slab center of a nonpolar slab. The surface is always *polar* if there is *no isometry of the form in equation 3*. All surfaces are polar for point group types 1 (space group type *P*1, number 1) and 3 (space group types *P*3, *P*3$_1$, *P*3$_2$, and *R*3, numbers 143-146).

A slab centered at a potential slab center would fall into one of the four categories: nonpolar type A, nonpolar type B, or stoichiometric when all atoms at the boundaries are included, nonpolar type C, or not nonpolar type B but stoichiometric after halving atoms on the boundaries, and non-stoichiometric. Variables *f* and *g* used in this section are integers. The *"unit layer thickness"* $z_{nu}$ is defined as the minimum $z_{nu}$ where a slab with boundaries $(z_{n-}, z_{n+}) = (z, z + z_{nu})$ is identical to one with $(z_{n-}, z_{n+}) = (z + z_{nu}, z + 2z_{nu})$ after any necessary in-plane rotation and/or translation. Figs. 2(c) and 2(d) are slabs of WAl$_2$ with the (001) surface where $(z_{1-}, z_{1+}) = (0, 1/3)$ and $(z_{1-}, z_{1+}) = (1/3, 2/3)$, respectively, as seen from the [001] direction. These two slabs contain one layer of WAl$_2$ each and are identical after in-plane translation, which shows that $z_{1u} = 1/3$ for this surface. The slab centers in Figs. 2(c) and 2(d) are $z_{1c} = 1/6$ and $z_{1c} = 1/2$, respectively.

$z_{nu}$ is the minimum positive $w_3$ ($0 < w_3 \leq 1/n$ must hold for an *n*-supercell) in an isometry of the form

$$(\boldsymbol{W}, \boldsymbol{w}) = \begin{pmatrix} \# & \# & \# & \# \\ \# & \# & \# & \# \\ 0 & 0 & 1 & w_3 \end{pmatrix}. \qquad (4)$$

When the isometry in equation 4 is applied as $\tilde{\boldsymbol{x}} = \boldsymbol{W}\boldsymbol{x} + \boldsymbol{w}$ when the basis vectors are $(\mathbf{a}_P, \mathbf{b}_P, \mathbf{c}_{nP})$, an atom with coordinate $x_3$ is mapped to a position with coordinate $\tilde{x}_3 = x_3 + w_3$, which means that the isometry in equation 1 maps all atoms, after necessary in-plane rotation and/or translation, to a position shifted by $w_3$ along $\mathbf{c}_{nP}$. It can be easily proved that $z_{1u}$ must be a reciprocal of an integer, in other words, can be written in the form $z_{1u} = 1/m$. Symmorphic space groups do not have isometries with intrinsic translation, hence $z_{1u} = 1$ always hold.

All potential slab centers can be derived once the unit layer thickness and one potential slab center is identified. This is because an isometry of the form

$$(\boldsymbol{W}, \boldsymbol{w}) = \begin{pmatrix} \# & \# & \# & \# \\ \# & \# & \# & \# \\ 0 & 0 & -1 & w_3 + fz_{nu} \end{pmatrix} \qquad (5)$$

must exist for any integer *f* if there is an isometry of the form

$$(\mathbf{W}, \mathbf{w}) = \begin{pmatrix} \# & \# & \# & \# \\ \# & \# & \# & \# \\ 0 & 0 & -1 & w_3 \end{pmatrix}. \tag{6}$$

The *unique potential slab center* $z_{Pc} = w_3/2$ is defined for the $(hkl)$ primitive cell using $w_3$ that makes equation 6 an isometry and satisfies $0 \leq w_3 < z_{1u}$. Therefore, using $z_{Pc}$ defined within the range $0 \leq z_{Pc} < z_{1u}/2$,

$$z_{1c} = f\frac{z_{1u}}{2} + z_{Pc} \tag{7}$$

must be a potential slab center. In the example of the WAl$_2$ (001) surface in Fig. 2, $z_{Pc} = 0$, $z_{1u} = 1/3$, and slab centers of the (001) 1-supercell have the form $f/6$. $z_{Pc}$ is always zero in symmorphic space groups because there are no isometries with an intrinsic translation part. The potential slab centers in an $(hkl)$ $n$-supercell can be derived from those in the $(hkl)$ primitive cell as

$$z_{nc} = \frac{z_{1c}}{n} = \left( f\frac{z_{1u}}{2} + z_{Pc} \right)\frac{1}{n}. \tag{8}$$

Fig. 5 shows (001) 1-, 2-, and 3-supercells of SnI$_4$ [31] (space group type $P23$, number 195) where $z_{nu} = 1/n$. One potential slab center is $z_{1c} = 0.5/1$ in the (001) 1-supercell, which relates to $z_{2c} = 0.5/2$ in the 2-supercell and $z_{3c} = 0.5/3$ in the 3-supercell.

Next, slab boundary candidates of a stoichiometric nonpolar slab are derived. If the polarity is determined for a given set of slab boundaries in an $(hkl)$ $n$-supercell, then slab boundaries resulting in a stoichiometric slab with the same polarity in an $(hkl)$ $m$-supercell can be identified based on the following two propositions. Here, the term "stoichiometric slab" is used to refer to a slab that is stoichiometric after any necessary surface reconstruction.

1) If $(z_{n-}, z_{n+})$ are slab boundaries of a stoichiometric slab, then $(z_{n-} + fz_{nu}, z_{n+} + gz_{nu})$ are slab boundaries of a stoichiometric slab with the same polarity.

2) If $(z_{n-}, z_{n+})$ are slab boundaries of a stoichiometric slab, then $(z_{m-}, z_{m+}) = ((m/n)z_{n-}, (m/n)z_{n+})$ are slab boundaries of a stoichiometric slab with the same polarity.

Fig. 5 can be used to intuitively understand these prepositions. Proposition 1 states that the slab boundary can be shifted by an integer multiple of $z_{nu}$, which indicates that a boundary at $z = 0.5/2$ in the 2-supercell can be shifted to $z = 1.5/2$ and the boundary at $z = 0.5/3$ in the 3-supercell can be shifted to $z = 1.5/3$ or $z = 2.5/3$ while retaining the same surface. Proposition 2 indicates that the same boundary can be transferred from an $(hkl)$ $n$-supercell to an $(hkl)$ $m$-supercell by simply changing the denominator of the boundary from $n$ to $m$. For instance, the boundaries at $z = 0.5/2$ and $z = 1.5/2$ in the (001) 2-supercell are equivalent to the boundaries at $z = 0.5/3$ and $z = 1.5/3$ in the (001) 3-supercell, respectively. For this reason, it is convenient to express the $z$-coordinate in an $(hkl)$ $n$-supercell as a fraction with denominator $n$ because the boundary is equivalent after adding an integer to the numerator and/or changing the denominator.

The above analysis means that the polarity of slabs with various thicknesses in an $(hkl)$ $n$-supercell can be derived by investigation of slabs in an $(hkl)$ 3-supercell with slab thickness $z_{3t} = z_{3u}/2 = (z_{1u}/2)/3$ or $z_{3t} = z_{3u} = z_{1u}/3$ and slab center $z_{3c} = (z_{Pc} + 1)/3$ or $z_{3c} = (z_{Pc} + z_{1u}/2 + 1)/3$. Use of the 3-supercell as well as addition of 1/3 when obtaining the slab center ensures that the two boundaries of the slab is contained between 0 and 1, or in other words, $z = 0$ is always in the vacuum region. Identifying whether a nonpolar slab is nonpolar type A or not is an easy task; looking at section planes of orientation $(hkl)$ that contain atoms, the surface is, by definition, nonpolar type A if all section planes have the same stoichiometry.

Three examples are provided to illustrate how $(hkl)$ 3-supercells are used to identify potential slab centers and the polarity. First, the (100) surface of BeSO$_4$ (Fig. 6 (a, b), space group type $I\bar{4}$, number 82) [32] is investigated. The space group type is symmorphic, hence $z_{1u} = 1$ and $z_{Pc} = 0$. The slab boundaries and the corresponding polarity are shown in Table 3. Next, the (011) surface of the high pressure phase of AgI (Fig. 6 (c, e), space group type *P*4/*nmm*, number 129) [33] is explored, where $z_{1u} = 1$ and $z_{Pc} = 0.25$. The slab boundaries and the corresponding polarity are shown in Table 4. Finally, the (010) surface of FeSe$_2$ (Fig. 6 (d, f), space group type *Pnnm*, number 58) [34] is analyzed, where $z_{1u} = 0.5$ and $z_{Pc} = 0$. The slab boundaries and the corresponding polarity are shown in Table 5.

The necessity of surface reconstruction forces an additional step in high-throughput calculations. This step of finding the appropriate reconstruction that is dependent on the crystal structure and surface orientation is a difficult and time-consuming task. Therefore, nonpolar slabs that do not need reconstruction, which are nonpolar type A and B surfaces, are desirable compared to nonpolar type C or polar surfaces. However, use of a simply cleaved slab may be inappropriate even in a nonpolar type B surface. For instance, if the SO$_4$ tetrahedra centered at $z = 1/3$ in Fig. 3(b) is slightly rotated such that the tetrahedra are not contained between the lines drawn at $z = 0.5/3$ and $z = 1.5/3$, then using $z = 0.5/3$ and $z = 1.5/3$ as slab boundaries would form a nonpolar slab but the SO$_4$ tetrahedra are not kept intact. In this case, it would be more appropriate to reconstruct the surface such that SO$_4$ tetrahedra are retained. This problem actually arises in BPO$_4$ [35] that is isostructural to BeSO$_4$. In another example, the nonpolar type A (110) surface of chalcopyrite CuGaSe$_2$ and CuInSe$_2$, which are important materials with photovoltaic applications, stabilizes by forming facets of (112) and (11$\bar{2}$) surfaces that each contain defects [36]. Another issue that requires attention is that all nonpolar slabs of the same surface orientation of the same crystal are not necessarily equivalent. In the (011) surface of AgI shown in Fig. 6(e), a slab cleaved at $(z_{3-}, z_{3+}) = (0.75/3, 1.75/3)$ and another at $(z_{3-}, z_{3+}) = (1.25/3, 2.25/3)$ are both nonpolar type B. However, the types of bonds that are severed in the two surfaces are different; therefore the surface energy would be different along with other properties that depend on the surface. In such a situation, investigation of various nonpolar slab terminations is necessary to find the most reasonable surface.

## 6. Automatic generation of nonpolar slabs

Assuming its existence, automatic generation of a nonpolar slab, where slab and vacuum thicknesses each exceed a given minimum thickness for any orientation of any crystal is desirable. Table 6 is a summary of relevant thicknesses in units of length. The basic flow is to determine the slab thickness based on the given minimum slab thickness and whether the slab thickness is to be an integer or half-integer multiple of the unit layer thickness. Next, the total cell thickness is derived such that the cell thickness is an integer multiple of the $(hkl)$ primitive cell thickness and the vacuum thickness is determined to be as small as possible while exceeding the minimum vacuum thickness.

Two examples of slab generation process are shown in Fig. 7. First, we build a (001) slab of $Li_2O_2$ (space group type $P6_3/mmc$, number 194, $c$ = 3.855 Å) [37] with a minimum slab thickness $t_{s0}$ of 9 Å and a minimum vacuum thickness $t_{v0}$ of 6 Å [Fig. 7(a)]. The thickness of the (001) primitive cell is $t_P = 3.855$ Å and the unit layer thickness is $t_u = t_P/2 = 1.9275$ Å. The thickness of a nonpolar (001) slab of $Li_2O_2$ must be an integer multiple of $t_u$, thus $t_s = 5t_u = 9.6375$ Å. The total slab thickness becomes $t_c = 5t_P = 19.275$ Å. Next, a (100) slab of massicot phase PbO (space group type $Pbcm$, number 57, $c$ = 4.743 Å) [38] with a minimum slab thickness $t_{s0}$ of 10 Å and a minimum vacuum thickness $t_{v0}$ of 10 Å is created [Fig. 7(b)]. Here, $t_u = t_P = 4.743$ Å. The thickness of the slab $t_{sh}$ will be a half-integer times $t_u$, namely $t_s = 2.5t_u = 11.8575$ Å, and the total slab thickness is $t_c = 5t_P = 23.715$ Å

The size of the generated $(hkl)$ $n$-supercell is $n = t_c/t_u$. The lower boundary of the slab is obtained based on the relevant value for the 3-supercell as $z_{n-} = (3z_{3-} - \lfloor 3z_{3-} \rfloor)/n$, which is 0 in both Figs. 7(a) and 7(b). Finally, one may wish to minimize the length of the out-of-plane basis vector by using $\mathbf{c}'_{nP} = \mathbf{c}_{nP} - p\mathbf{a}_P - q\mathbf{b}_P$ instead of $\mathbf{c}_{nP}$. The vector $\mathbf{c}'_{nP}$ is shortest, that is, maximally orthogonal to $\mathbf{a}_P$ and $\mathbf{b}_P$, when $p$ and $q$ are integers closest to $(\mathbf{c}_{nP} \cdot \mathbf{a}_P)/|\mathbf{a}_P|^2$ and $(\mathbf{c}_{nP} \cdot \mathbf{b}_P)/|\mathbf{b}_P|^2$, respectively.

## 7. Summary

We categorize surface polarity based on a crystallographic approach. Polar surfaces are defined as surfaces that do not have an isometry in the form of equation 3 when basis vectors **a** and **b** are in-plane vectors. Surfaces that are not polar are further categorized into three types. Every layer of atoms is stoichiometric in a nonpolar type A surface. Other surfaces are either nonpolar type B or nonpolar type C. The surface is the former if the boundaries of the repeat unit with no macroscopic dipole moment perpendicular to the surface lie between layers of atoms. A nonpolar type C slab cannot be nonpolar and stoichiometric at the same time when a slab is simply cleaved from bulk, but a nonpolar and stoichiometric slab can be obtained with reconstruction of the surface, for example, by removing half of the atoms on the topmost layer on both sides. Moreover, this study outlines a procedure using isometries that identifies whether stoichiometric nonpolar slabs can be obtained for any surface orientation of any crystal, as summarized in Fig. 8. This procedure also derives the slab boundary positions in the out-of-plane direction of nonpolar slabs as well as the polarity type. Automatic generation of nonpolar slabs will certainly be a powerful tool in exploration of surface phenomena and properties, and would be a necessity rather than a convenience in high-throughput studies where many surfaces have to be generated and investigated.

**Acknowledgments**

This study was supported by a Grant-in-Aid for Scientific Research on Innovative Areas "Nano Informatics" (grant number 25106005) and the MEXT Elements Strategy Initiative to Form Core Research Center. The VESTA code was used to draw Figs. 1-3 and 5-7. [39]

**Table 1.** Definition of $\mathbf{M'}$ to obtain $(\mathbf{a'}, \mathbf{b'}, \mathbf{c'})$.

| Two zeros in $\{h,k,l\}$ | | | One zero in $\{h,k,l\}$ | | | No zeros in $\{h,k,l\}$ |
|---|---|---|---|---|---|---|
| $h \neq 0$ | $k \neq 0$ | $l \neq 0$ | $h = 0$ | $k = 0$ | $l = 0$ | |
| $\begin{pmatrix} 0 & 0 & 1 \\ 1 & 0 & 0 \\ 0 & 1 & 0 \end{pmatrix}$ | $\begin{pmatrix} 0 & 1 & 0 \\ 0 & 0 & 1 \\ 1 & 0 & 0 \end{pmatrix}$ | $\begin{pmatrix} 1 & 0 & 0 \\ 0 & 1 & 0 \\ 0 & 0 & 1 \end{pmatrix}$ | $\begin{pmatrix} 1 & 0 & 0 \\ 0 & l & k \\ 0 & -k & l \end{pmatrix}$ | $\begin{pmatrix} 0 & l & h \\ 1 & 0 & 0 \\ 0 & -h & l \end{pmatrix}$ | $\begin{pmatrix} 0 & k & h \\ 0 & -h & k \\ 1 & 0 & 0 \end{pmatrix}$ | $\begin{pmatrix} k & l & h \\ -h & 0 & k \\ 0 & -h & l \end{pmatrix}$ |

**Table 2.** Basis vectors $(\mathbf{a}_P, \mathbf{b}_P, \mathbf{c}_P)$ of $(hkl)$ primitive cells in Cartesian coordinates when the conventional cell is $(\mathbf{a}, \mathbf{b}, \mathbf{c}) = \begin{pmatrix} a & 0 & 0 \\ 0 & a & 0 \\ 0 & 0 & a \end{pmatrix}$.

| Bravais lattice | $(hkl)$ primitive cell | | | |
|---|---|---|---|---|
| | $(100)$ | $(110)$ | $(111)$ | $(211)$ |
| Simple cubic | $\begin{pmatrix} 0 & 0 & a \\ a & 0 & 0 \\ 0 & a & 0 \end{pmatrix}$ | $\begin{pmatrix} 0 & a & a \\ 0 & \bar{a} & 0 \\ a & 0 & 0 \end{pmatrix}$ | $\begin{pmatrix} a & a & a \\ \bar{a} & 0 & 0 \\ 0 & \bar{a} & 0 \end{pmatrix}$ | $\begin{pmatrix} 0 & a & a \\ \bar{a} & \bar{a} & \bar{a} \\ a & \bar{a} & 0 \end{pmatrix}$ |
| Face-centered cubic | $\begin{pmatrix} 0 & 0 & \frac{a}{2} \\ \frac{a}{2} & \frac{\bar{a}}{2} & 0 \\ \frac{a}{2} & \frac{a}{2} & \frac{a}{2} \end{pmatrix}$ | $\begin{pmatrix} \frac{a}{2} & 0 & \frac{a}{2} \\ \frac{\bar{a}}{2} & 0 & 0 \\ 0 & \bar{a} & \frac{\bar{a}}{2} \end{pmatrix}$ | $\begin{pmatrix} \frac{a}{2} & \frac{a}{2} & a \\ \frac{\bar{a}}{2} & 0 & 0 \\ 0 & \frac{\bar{a}}{2} & 0 \end{pmatrix}$ | $\begin{pmatrix} 0 & a & \frac{a}{2} \\ \frac{\bar{a}}{2} & \bar{a} & \frac{\bar{a}}{2} \\ \frac{a}{2} & \bar{a} & 0 \end{pmatrix}$ |
| Body-centered cubic | $\begin{pmatrix} 0 & 0 & \frac{a}{2} \\ a & 0 & \frac{a}{2} \\ 0 & a & \frac{a}{2} \end{pmatrix}$ | $\begin{pmatrix} \frac{a}{2} & 0 & \frac{a}{2} \\ \frac{\bar{a}}{2} & 0 & \frac{a}{2} \\ \frac{a}{2} & \bar{a} & \frac{\bar{a}}{2} \end{pmatrix}$ | $\begin{pmatrix} a & a & \frac{3a}{2} \\ \bar{a} & 0 & \frac{\bar{a}}{2} \\ 0 & \bar{a} & \frac{\bar{a}}{2} \end{pmatrix}$ | $\begin{pmatrix} \frac{a}{2} & 0 & \frac{a}{2} \\ \frac{\bar{a}}{2} & a & \frac{a}{2} \\ \frac{\bar{a}}{2} & \bar{a} & \frac{\bar{a}}{2} \end{pmatrix}$ |

**Table 3.** Slab boundaries investigated in the (100) slab of $BeSO_4$.

| $3z_{3t}$ | $3z_{3c}$ | $3z_{3-}$ | $3z_{3+}$ | Polarity |
|---|---|---|---|---|
| 0.5 | 1 | 0.75 | 1.25 | Non-stoichiometric |
| 0.5 | 1.5 | 1.25 | 1.75 | Non-stoichiometric |
| 1 | 1 | 0.5 | 1.5 | Nonpolar type B |
| 1 | 1.5 | 1 | 2 | Nonpolar type C |

**Table 4.** Slab boundaries investigated in the (011) slab of high pressure phase AgI.

| $3z_{3t}$ | $3z_{3c}$ | $3z_{3-}$ | $3z_{3+}$ | Polarity |
|---|---|---|---|---|
| 0.5 | 1.25 | 1 | 1.5 | Nonpolar type B |
| 0.5 | 1.25 | 0.75 | 1.75 | Nonpolar type B |
| 1 | 1.75 | 1.5 | 2 | Non-stoichiometric |
| 1 | 1.75 | 1.25 | 2.25 | Nonpolar type B |

**Table 5.** Slab boundaries investigated in the (010) slab of $FeSe_2$.

| $3z_{3t}$ | $3z_{3c}$ | $3z_{3-}$ | $3z_{3+}$ | Polarity |
|---|---|---|---|---|
| 0.25 | 1 | 0.875 | 1.125 | Non-stoichiometric |
| 0.25 | 1.25 | 1.125 | 1.375 | Non-stoichiometric |
| 0.5 | 1 | 0.75 | 1.25 | Nonpolar type B |
| 0.5 | 1.25 | 1 | 1.5 | Nonpolar type C |

**Table 6.** Definition of thicknesses relevant to automatic slab generation.

| Thickness | Symbol |
|---|---|
| Minimum slab thickness (given) | $t_{s0}$ |
| Minimum vacuum thickness (given) | $t_{v0}$ |
| Thickness of the $(hkl)$ primitive cell | $t_P = \dfrac{(\mathbf{a}_P \times \mathbf{b}_P) \cdot \mathbf{c}_{1P}}{|\mathbf{a}_P \times \mathbf{b}_P|}$ |
| Unit layer thickness | $t_u = z_{1u} t_P$ |
| Slab thickness if integer multiple of $t_u$ | $t_s = \left\lceil \dfrac{t_{s0}}{t_u} \right\rceil t_u$ |
| Slab thickness if half-integer multiple of $t_u$ | $t_s = \left( \left\lceil \dfrac{t_{s0}}{t_u} - \dfrac{1}{2} \right\rceil + \dfrac{1}{2} \right) t_u$ |
| Total cell thickness | $t_c = \left\lceil \dfrac{t_s + t_{v0}}{t_P} \right\rceil t_P$ |

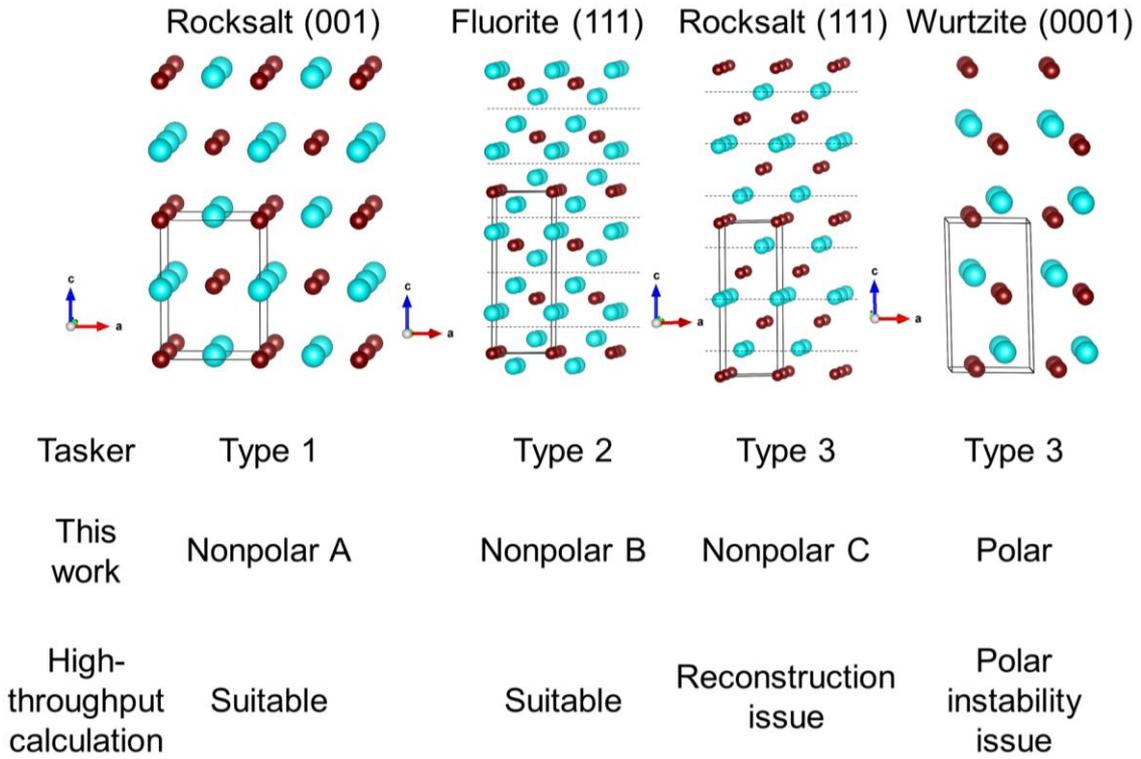

**Fig. 1.** Definition of slab polarity types based on Tasker [7] and this work as well as issues in high-throughput calculations. Small brown circles and large blue circles indicate cations and anions, respectively. The unit cells shown are the smallest supercell that is primitive in-plane and $\mathbf{a}\cdot\mathbf{c}=\mathbf{b}\cdot\mathbf{c}=0$. Dotted lines in fluorite (111) and rocksalt (111) indicate the repeat unit where the macroscopic dipole moment perpendicular to the surface is absent in a slab of this repeat unit. One can alternatively choose to place the repeat unit boundaries on cation planes instead of anion planes in rocksalt (111).

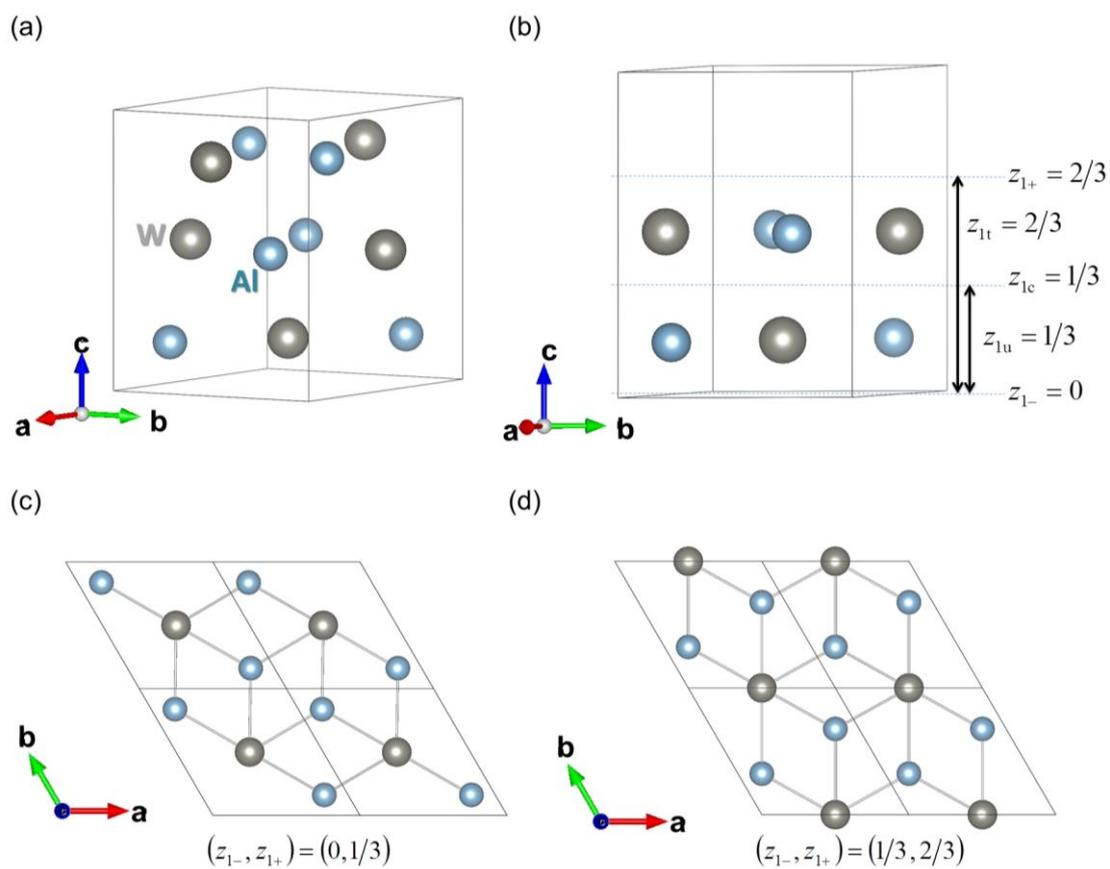

**Fig. 2.** (a) (001) 1-supercell (primitive cell) of WAl$_2$ and (b-d) (001) slab of WAl$_2$ where (b) $(z_{1-}, z_{1+}) = (0, 2/3)$, (c) $(z_{1-}, z_{1+}) = (0, 1/3)$, and (d) $(z_{1-}, z_{1+}) = (1/3, 2/3)$. Al-W bonds are shown in (c) and (d) for clarity.

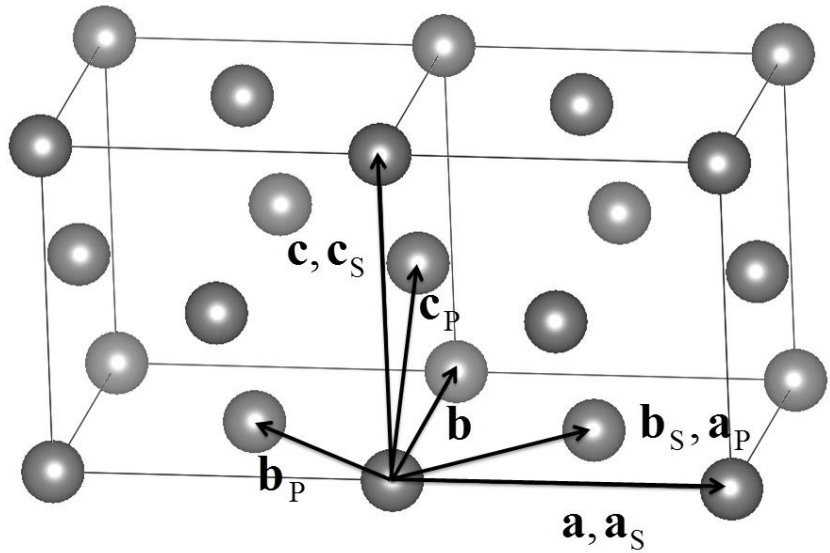

**Fig 3.** Various basis vectors relevant to the (001) surface of a face-centered cubic cell. The boxes indicate the conventional cell. $(\mathbf{a},\mathbf{b},\mathbf{c})$ and $(\mathbf{a}_P,\mathbf{b}_P,\mathbf{c}_{1P})$ are basis vectors of the conventional cell and (001) primitive cell, respectively. $(\mathbf{a}_S,\mathbf{b}_S,\mathbf{c}_S)$ are basis vectors that appear during the derivation of $(\mathbf{a}_P,\mathbf{b}_P,\mathbf{c}_{1P})$.

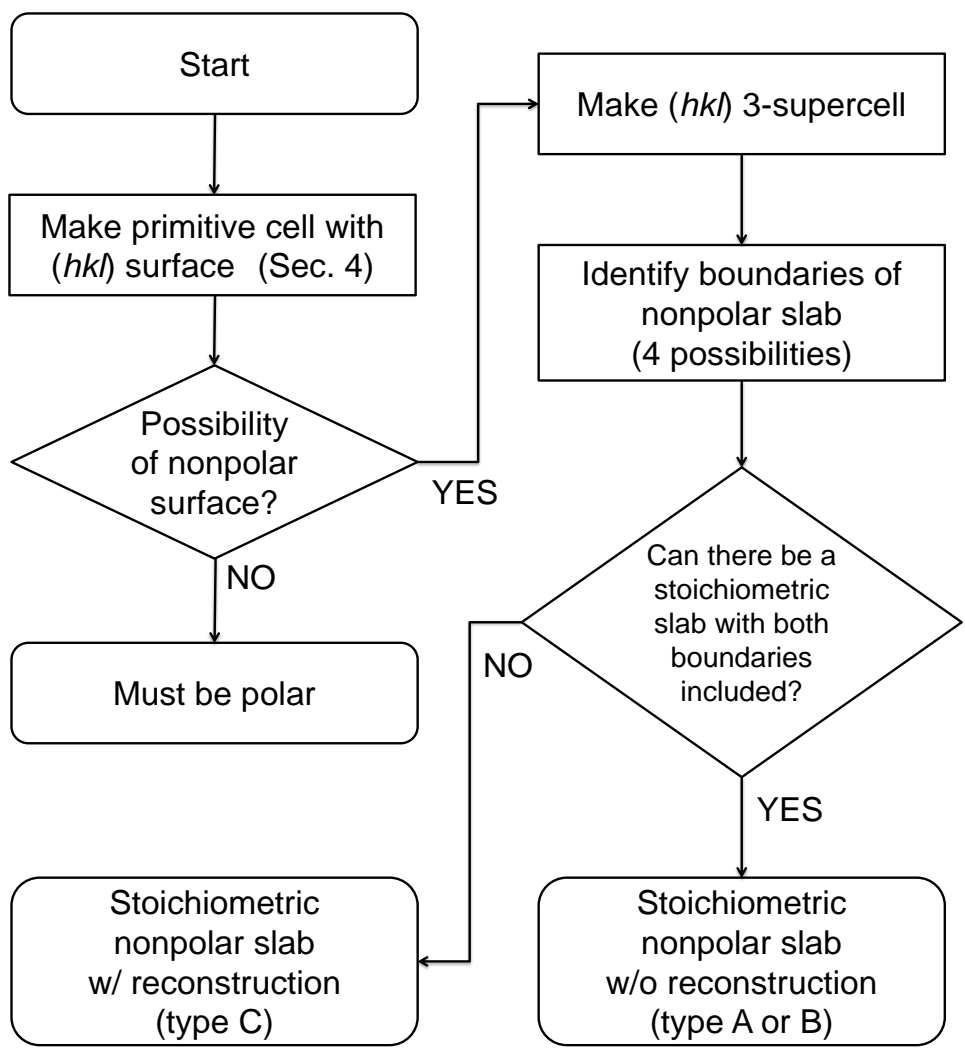

**Fig. 4.** Simple flowchart of the algorithm described in Section 5.

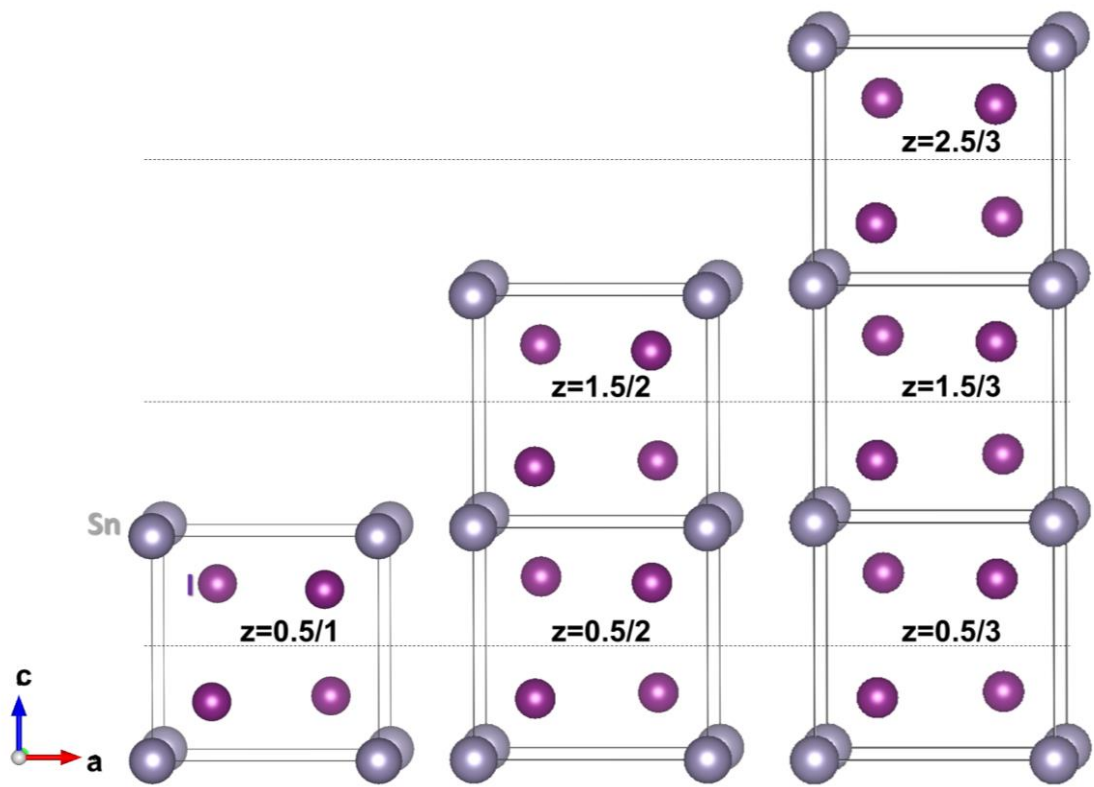

**Fig. 5**. (001) 1-, 2-, and 3-supercells of $SnI_4$.

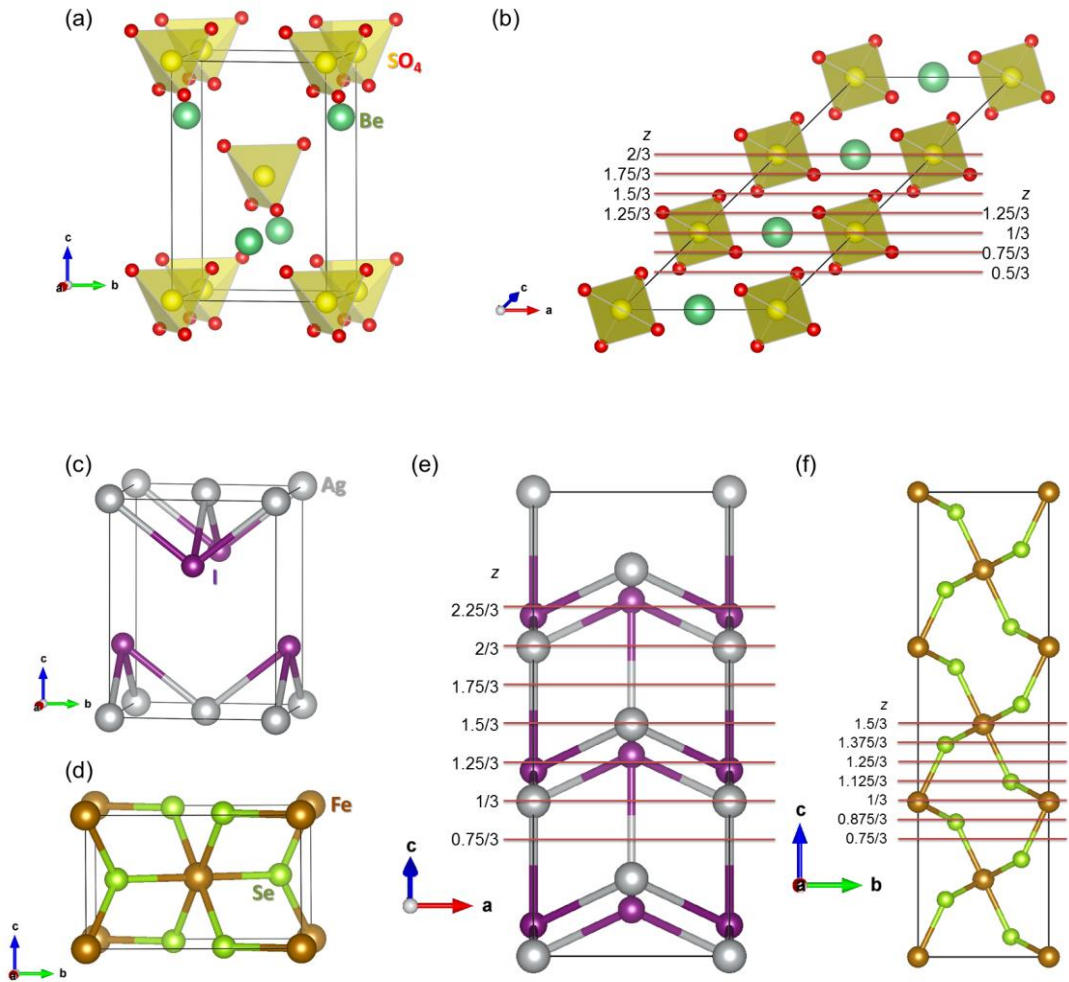

**Fig. 6.** Examples of nonpolar slab identification. (a) Unit cell and (b) (100) 3-supercell of BeSO$_4$. Unit cells of (c) high pressure phase AgI and (d) FeSe$_2$, and (e) AgI (011) and (f) FeSe$_2$ (010) 3-supercells. Legend: BeSO$_4$: green circles indicate Be and SO$_4$ are shown as tetrahedra. AgI: Large gray circles are Ag while small purple circles are I. FeSe$_2$: Large brown circles are Fe and small green circles are Se. Lines in 3-supercells are drawn at $z = 0.25/3$ intervals in BeSO$_4$ and AgI and $z = 0.125/3$ intervals in FeSe$_2$.

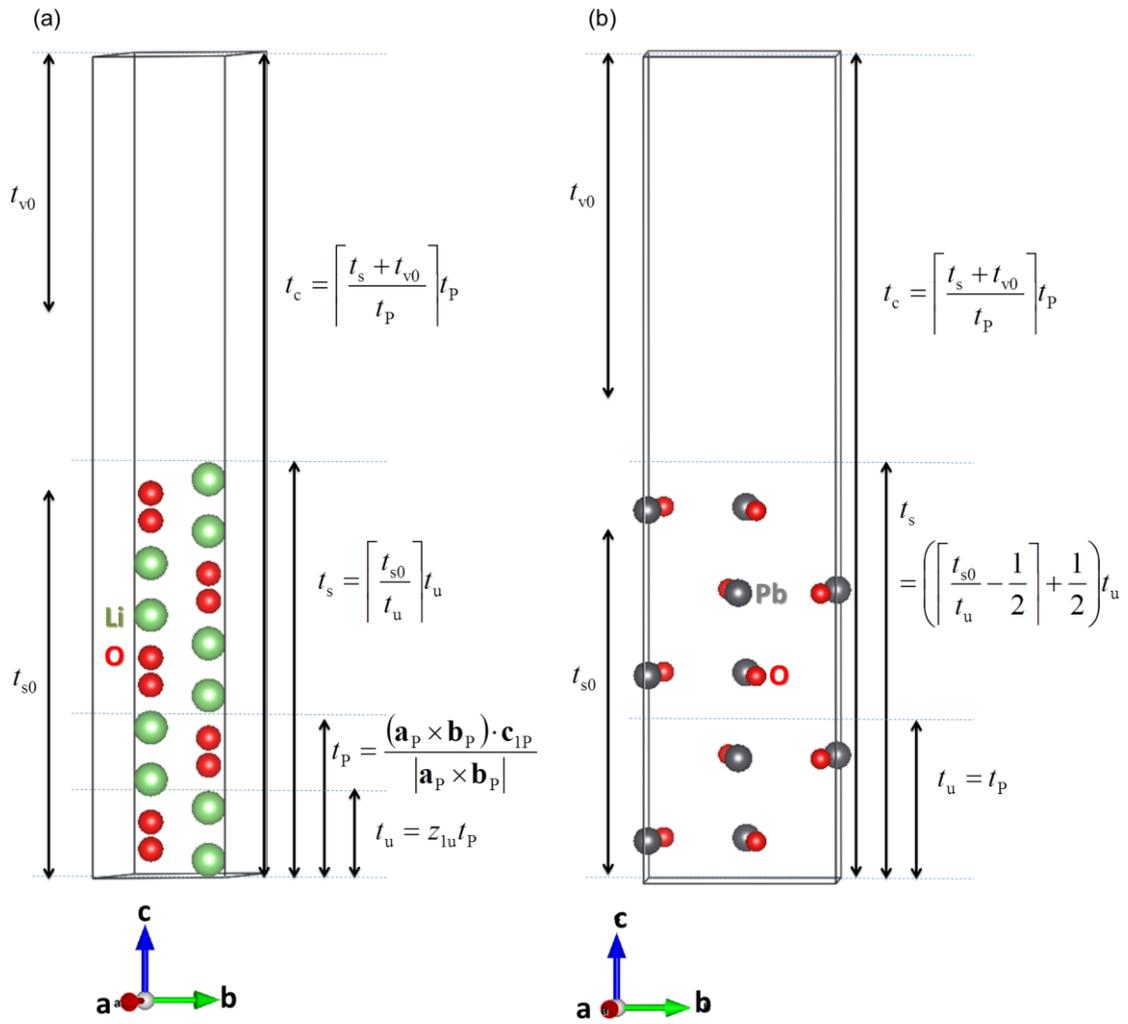

**Fig. 7.** Generation of (a) a $Li_2O_2$ (001) slab and (b) a massicot phase PbO (001) slab.

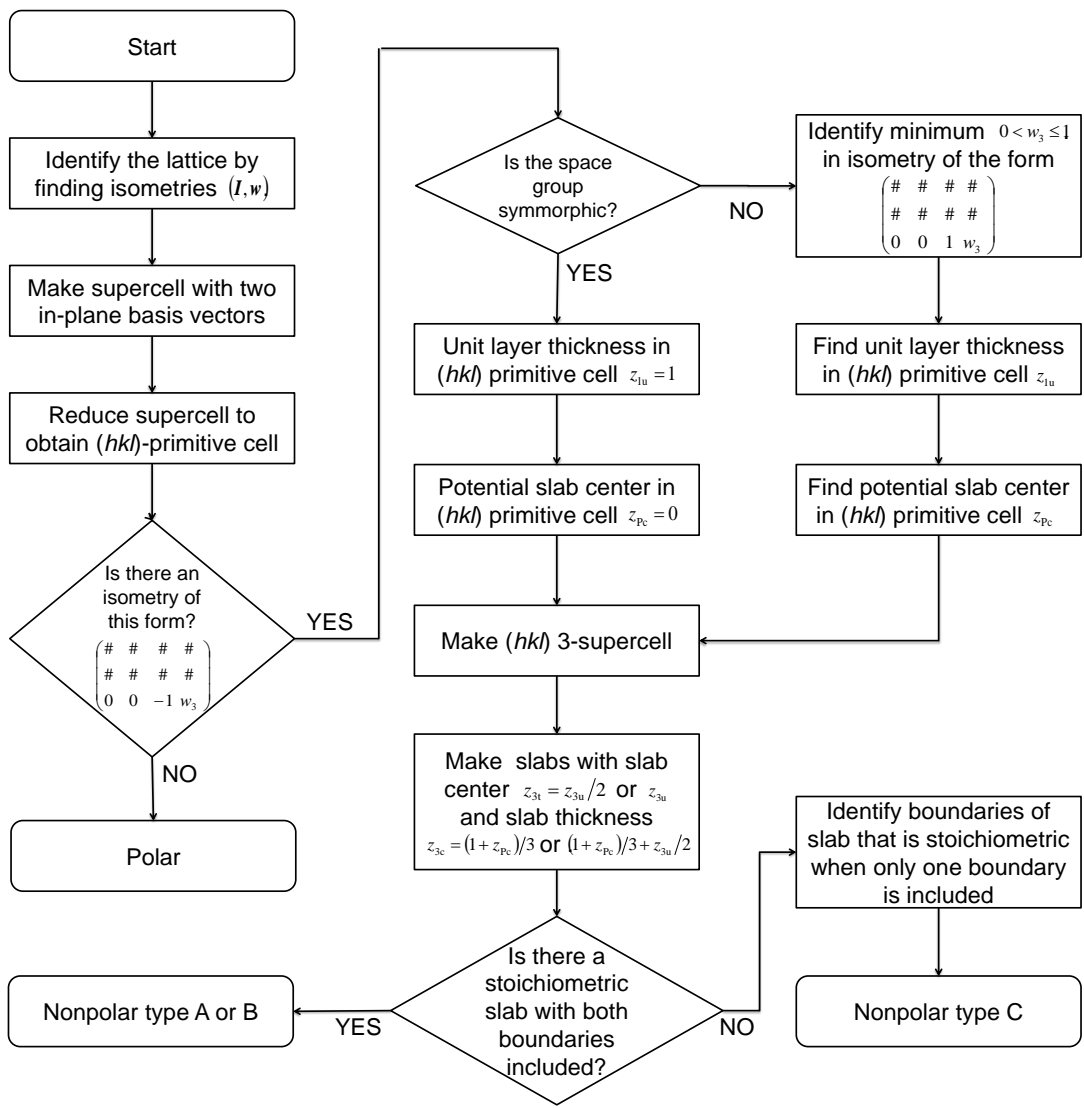

**Fig. 8.** Detailed flowchart of the proposed algorithm to identify nonpolar slabs.

**Appendix A. Two-dimensional Gaussian lattice reduction**

The following Gaussian lattice reduction algorithm finds the two non-zero shortest lattice vectors in a plane that are not linearly dependent (adapted from [40]). Vectors $u$ and $v$ are reduced through the following algorithm:

1) If $u \cdot v < 0$, $-v \to v$.

2) If $|u| < |v|$, $(v, u) \to (u, v)$.

3) Repeat $\left( v, u - \left\lfloor \dfrac{u \cdot v}{|v|^2} \right\rfloor v \right) \to (u, v)$ until $|u| \leq |v|$.

4) If $\begin{cases} u \cdot v < |v|^2/2, (v, u) \to (u, v) \\ |u - v| > |v|, (v, u - v) \to (u, v) \\ \text{otherwise}, (u - v, -v) \to (u, v) \end{cases}$.